\documentstyle[11pt,aaspp4]{article}

\slugcomment{To appear in Astronomical Journal}

\lefthead{S.C. Keller}
\righthead{IR photometry of Young MC Clusters}

\begin{document}

   \title{Infrared Photometry of Red Supergiants in Young Clusters in the Magellanic Clouds}

\author{Stefan C. Keller}
\affil{Research School of Astronomy and Astrophysics, The Australian National
University, Weston Creek P.O., ACT 2611, Australia.}
 
\begin{abstract} 
We present broad-band infrared photometry for 52 late-type supergiants in the young Magellanic Clouds clusters NGC 330, NGC 1818, NGC 2004 and NGC 2100. Standard models are seen to differ in the temperature they predict for the red supergiant population on the order of 300K. It appears that these differences most probably due to the calibration of the mixing-length parameter, $\alpha_{P}$, in the outermost layers of the stellar envelope. Due to the apparent model dependent nature of $\alpha_{P}$ we do not quantitatively compare $\alpha_{P}$ between models. Qualitatively, we find that $\alpha_{P}$ decreases with increased stellar mass within standard models. We do not find evidence for a metallicity dependence of $\alpha_{P}$.

\end{abstract}

 
\section{Introduction}
The young clusters NGC 330 in the SMC and NGC 1818, NGC 2004 and NGC 2100 have
been the focus of several previous studies aimed at understanding the
evolution of intermediate mass stars in a low metallicity environment. These
clusters are standout targets for such studies given their relative
richness. They present a significant sample of stars in evolved blue and red
supergiant phases.

A thorough examination of the evolution of intermediate mass stars requires a matching of all pertinent features of the observed CMD. The location of the red supergiant population on the CMD is a fundamental input into any such study. We have obtained HST/WFPC2 photometry in the visible and far-UV for these four clusters (Keller et al \cite{evolution}). With this data we will closely examine the compatibility between our observations and the predictions of standard evolutionary models in a future paper. It is the goal of the present work to obtain accurate temperatures and luminosities for the red supergiant (RSG) population through an examination of their IR colours.

Previous IR studies within these clusters have been limited. A sample of the
outlying RSGs from NGC 330, 2004 and 2100 were observed by McGregor \& Hyland
(\cite{mcg84}). A similar sample was also taken by \cite{car85} in NGC 330. These studies drew upon the derived temperatures and luminosities of these RSGs as evidence for a low metal abundance in NGC 330. Subsequent discussion has relied upon these IR colours and the $B$$-$$V$ colours of Robertson (\cite{rob74}). 

The observed properties of the RSG population has featured prominently in many
of the subsequent discussions of the evolutionary implications of the
populations of these clusters. Many of these discussions have to some extent
been superseded with the introduction of new opacities and updated input
physics. The first detailed approach to constraining the stellar evolutionary
models for intermediate mass stars was made by Stothers and Chin (\cite{sto92}) in their study of NGC 330. They set out to constrain the criterion for convection and semi convection and the degree of convective core overshoot. They found that only with the implementation of the Ledoux criterion could RSGs be produced at low metallicities. They used the luminosity difference between the blue and RSGs to rule out significant amounts of convective core overshoot (d/Hp$<$0.2). In the study of NGC 330 by Chiosi et al (\cite{chi95}) it was shown that with new opacities and input physics it was possible to produce blue supergiants with the Schwarzschild criterion at low metallicity. Their models were unable to match both the MS turnoff and the RSG luminosities simultaneously; they took this to indicate that there was a substantial age spread within the cluster. In the study by Caloi and Cassatella (\cite{cal95}) of NGC 2004 attention was drawn to the fact that the RSGs appeared underluminous relative to the upper MS. This result, it was acknowledged, was not definitive and required the acquisition of more accurate temperatures for the cluster RSGs.

The RSGs of the present study provide a useful calibration point for the degree of convection in stellar envelopes. The standard treatment is within the mixing-length theory (see for example Bohm-Vitense \cite{boh58}). This contains the average eddy mixing length parameter, $\alpha_{P}$, which is varied to achieve a match between observed effective temperatures and those predicted by stellar models. The precise value of $\alpha_{P}$ is different from model to model; however it is typically treated as a constant over a wide range of masses, luminosities and metallicities.  Amongst low mass stars there are no signs of a dependence of $\alpha_{P}$ on stellar mass. The study of Stothers and Chin (\cite{sto95}) of galactic red giants and red supergiants in the mass range of 5-13 $M_{\odot}$ has shown that there is a significant decrease in $\alpha_{P}$ with increased mass. They sought to extend this analysis to the lower metallicity regime available within NGC 330 to investigate the trend of $\alpha_{P}$ with metallicity (Stothers and Chin \cite{sto96}). This study was inconclusive in this regard due to the uncertainty introduced by the degree of reddening to the cluster which has been a source of much debate. We claim that current determination of the reddening to NGC 330 is more convergent than discussed in Stothers and Chin (\cite{sto96}). We qualitatively discuss some of the implications for trends of $\alpha_{P}$ with metallicity and stellar mass; however, a quantitative study of $\alpha_{P}$ is beyond the scope of the current work and as we shall see may not be possible with present stellar evolutionary models.

\section{Observations and Data Reduction}
\subsection{IR photometry}
Broad band $J$, $K$ photometry was obtained on the MSSSO 2.3m telescope at
Siding Spring Observatory on 14-16 December, 1997 using the CASPIR
256$\times$256 pixel InSb array (see McGregor \cite{mcg94a} for more details on the instrument). Using a dithered pattern we have imaged the central 3.5\arcmin$\,$ of each cluster. Nearby photometric standard stars were observed frequently
interspersed with the cluster exposures. The standards HD20223 and HD52467 were
used for the LMC clusters and HD 1274 for NGC 330 in the SMC. The standard
values for these stars were taken from Carter \& Meadows (\cite{car95}),
transformation from the Carter SAAO system to the AAO system were made using
the transformations detailed in McGregor (\cite{mcg94b}). The results presented herein are on the AAO system, which is virtually identical to the Johnson system (Jones \& Hyland \cite{jon80}). Our results are presented in tables~\ref{tab1} -~\ref{tab4}.

If we compare our $J$,$K$ photometry of NGC 330 with that of \cite{car85}(CJF) we find $(J-K)=(J-K)_{CJF}+(0.06\pm0.08)$ and $K=K_{CJF}+(0.00\pm0.09)$. We consider the uncertainty within our $J$ and $K$ photometry to be $\pm0.05$mag in both bands.

\placetable{tab1}
\placetable{tab2}
\placetable{tab3}
\placetable{tab4}

\subsection{Optical Photometry}
These clusters are spatially dense; consequently, crowding is a serious problem
for those stars near the cluster centre. In previous studies it has not been
possible to use the $V$$-$$K$ colour of the red supergiants because of the uncertain effects of crowding. In the present study we utilise the HST/WFPC2 photometry of Keller et al (\cite{kel99b}) of those stars within the aforementioned field. Crowding is not a major concern in the WFPC2 fields. The F555W magnitudes have been converted to Johnson $V$ by utilising the transformation of Holtzman et al (\cite{hol95b}). For those stars in the outer extremities of the clusters we have relied upon the $V$  photometry of Keller et al. (\cite{kel99a}). We consider the uncertainty of our $V$ photometry from both sources to be $\pm$0.02mag for the objects considered in present work.

\section{Reddening corrections}

The adopted reddening corrections are taken from Keller et al. (\cite{kel99b}), namely E($B$$-$$V$)=0.08 for NGC 330, 1818, and 2004 and 0.24 for NGC 2100. Reddening ratios are taken from Bessell et al. (\cite{bes98}). The reddening assumed to each cluster is critical to the derived effective temperatures (T$_{eff}$). For instance a change of $\Delta$E($B$$-$$V$)=0.02 corresponds to a $\Delta$ T$_{eff}$=100K. 

The reddening to NGC 330 has been the focus of much debate, in particular the effect that the applied reddening has upon the derived metallicity from spectroscopic analysis of the cluster RSGs. Abundance studies provide an estimate of the temperature of the stars under analysis. However, the precision of such determinations is not high. It is commonly the case that recourse is made to the photometric colours of the program star in order to constrain the T$_{eff}$ used in the analysis. In this way the T$_{eff}$ of the target star and hence the resultant abundances are bound to the E($B$$-$$V$).

In the case of NGC 330 a low value of reddening (E($B$$-$$V$)=0.03) was found by \cite{car85} and Robertson (\cite{rob74}) from the $B$$-$$V$ colours for the blue supergiants within the cluster. Using this value of reddening a very low metallicity of [Fe/H]=-1.3 was found (Barbuy et al. (\cite{bar91}), Spite, Richtler and Spite (\cite{spi91}) and Meliani, Burbuy and Perrin (\cite{mel95})). This is a factor of three less abundant than the surrounding field - a curious finding. An examination of the colours of B type stars in the vicinity of the MS reveals a considerably higher value: E($B$$-$$V$)=0.12 (Bessell \cite{bes91}) or 0.09 (Caloi et al. \cite{cal93}). Hill (\cite{hil99a}) uses a E($B$$-$$V$)=0.09 and the T$_{eff}$ of the present work to derive a [Fe/H]=-0.88, which is on the order of that seen in the field. The recent study by Gonzalez \& Wallerstein (\cite{gon99}) find a similar metallicity of [Fe/H]=-0.94 solely from spectroscopic determinations of effective temperature. The convergence of the photometric and spectroscopic temperatures provides added confidence that value of reddening used here is sufficent.

\begin{figure}
\plotone{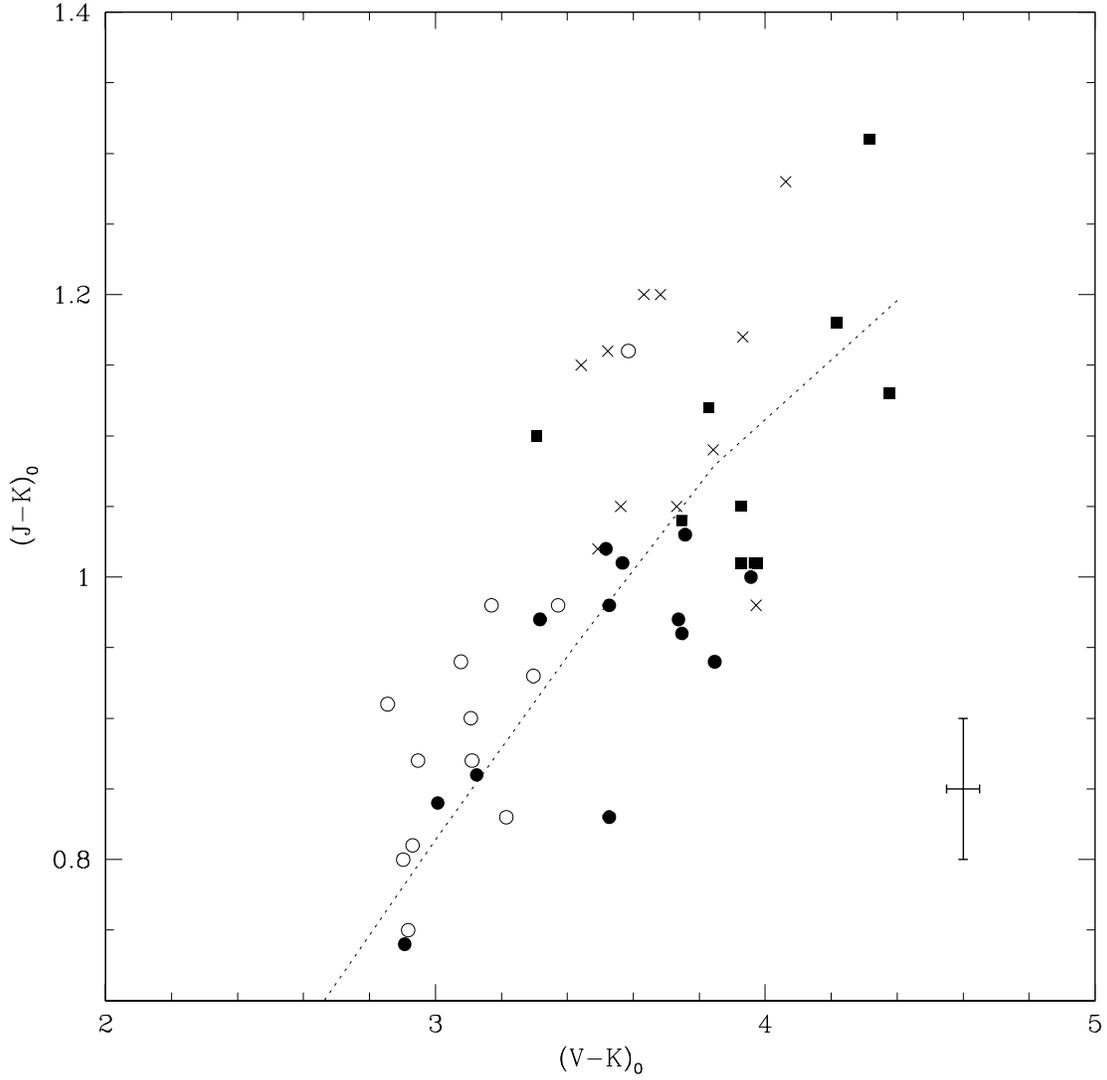}
\caption{J-K data plotted against V-K for the red supergiants in the four
clusters. The RSG of NGC 330 are shown by open circles, NGC 1818 by filled circles, NGC 2004 by solid boxes and NGC 2100 by crosses. The colours have been dereddened (see text for details). Approximate error bars are shown. The dashed line shows the predicted locus for logg=+0.5 and [Fe/H]=-0.60 (Plez et al. 1997).}
\label{vmkjmk}
\end{figure}

A colour-colour diagram of the de-reddened $V$$-$$K$ against $J$$-$$K$ is shown in figure~\ref{vmkjmk}. Also shown in figure~\ref{vmkjmk} is the predicted locus for logg=+0.5 and [Fe/H]=-0.60 (Plez et al. \cite{ple97}). This value for metallicity is intermediate between SMC and LMC metallicities. There are no radical departures from this relation. 

\placefigure{vmkjmk}
\placefigure{rsg}

\section{Effective Temperatures and Luminosities for the RSG Population}
\begin{figure}
\plotone{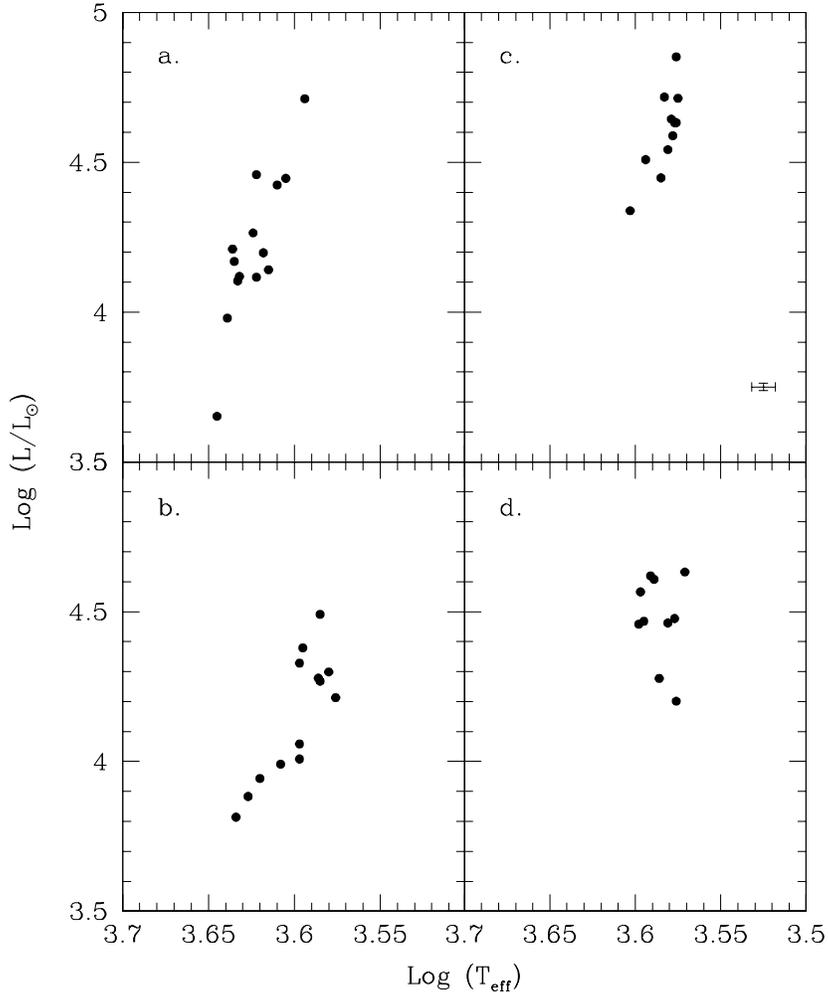}
\caption{H-R diagrams for the giant-branch members within (a) SMC cluster
NGC 330, and the LMC clusters (b) NGC 1818, (c) NGC 2004 and
(d) NGC 2100. Typical error bars are shown in figure~\ref{rsg}c.}
\label{rsg}
\end{figure}

The dereddened $V$$-$$K$ colours were converted into effective temperatures through the tables of Bessell et al. (\cite{bes98}). The bolometric correction required was similarly obtained from this source. Figures~\ref{rsg}a-d show the resultant H-R diagram for the cluster red supergiants. Figure~\ref{rsg}c shows appropriate error bars derived from the photometric uncertainties discussed above.

Recently Oliva and Origlia (\cite{oo98}) have examined the IR spectra of several young MC clusters, in particular they have focused on the strength of the CO(6,3) (1.62$\mu$m) band-head. They conclude that the red supergiants within these clusters must be cooler than predicted by evolutionary models in order to match the observed metallicities. They suggest that the discrepancy arises from the calibration of the mixing-length parameter, $\alpha_{P}$. However, the situation is not as discrepant as Oliva and Origlia have suggested. In their examination of NGC 330 they have required a fit to a metallicity of [Fe/H]=-1.3, which as discussed above, is much less than that suggested by recent determinations. In order to achieve a match between the predicted and observed IR spectrum, they require the red supergiant population to be of a temperature of 3600K. Figure~\ref{rsg}a rules this out to at least a 4 sigma level, instead the RSG within NGC 330 are distinctly hotter at an average temperature of $\sim 4100$K. Indeed, equivalent width of the CO(6,3) feature within NGC 330 is compatible with the predictions of standard models with [Fe/H]=-0.7 similar to that favoured by recent determinations (see Oliva and Origlia (\cite{oo98}) figure 3).

As discussed above, the study of Caloi and Cassatella (\cite{cal95}) found the RSGs to be of lower luminosity than the luminous tip of the MS. The temperatures and luminosities in Caloi and Cassatella (\cite{cal95}) are based upon previous $BV$ photometry and spectral classifications (see references therein). Caloi and Cassatella report that the most luminous blue supergiants within NGC 2004 have a log(L/L$_{\odot})$=4.9, whilst the most luminous RSG are at log(L/L$_{\odot})$=4.6 and logT$_{eff}$=3.62. Our observations indicate a mean temperature for the RSG clump some 400K cooler at logT$_{eff}$=3.58. The resultant difference in bolometric correction shifts the luminosity of the brightest RSG to log(L/L$_{\odot})$=4.83 and to greater concordance with the luminosities of the blue supergiants.

\placefigure{combined}
\placefigure{combinedgeneva}

\subsection{The Mixing Length Parameter - $\alpha_{P}$}

\begin{figure}
\plotone{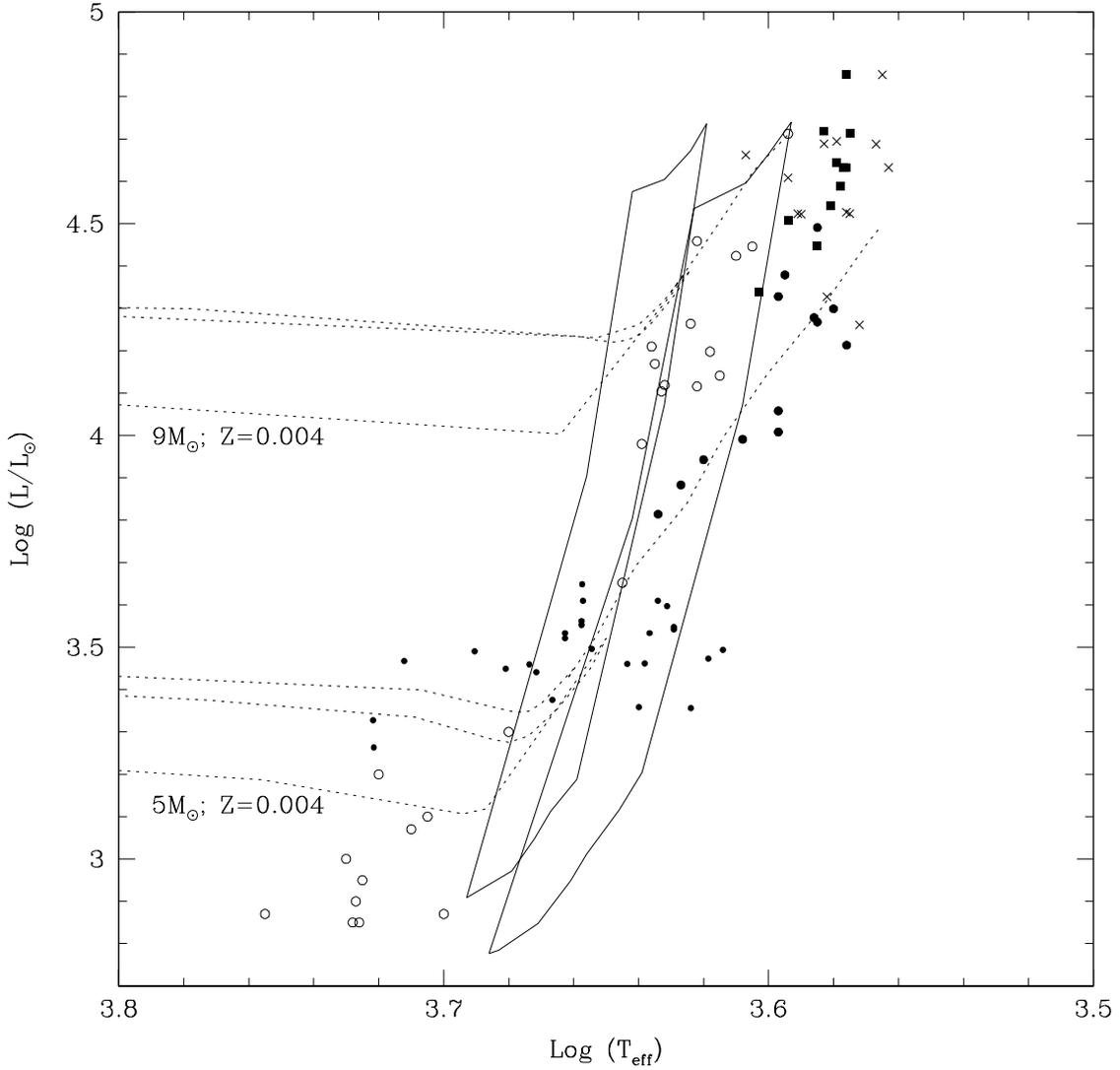}
\caption{H-R diagram showing evolutionary tracks from the models of Fagotto (\cite{fag94}) for Z=0.004 (dashed lines) at 9 and 5M$_{\odot}$. The solid lines contain the region of the downward slope of the evolutionary path in which the RSG spend the majority of their time for Z=0.004 (left) and Z=0.008 (right). The symbols described in figure~\ref{vmkjmk} show the RSG of the present work. Also shown at lower luminosities is the data for NGC 458 (open circles) and NGC 1850 (small filled circles).}
\label{combined}
\end{figure}

We show in figure~\ref{combined} the combined data from the four clusters, overlaid by the evolutionary tracks from Fagotto et al (\cite{fag94}) for [Fe/H]=-0.40 and -0.70 . We immediately see the separation between the lower metallicity NGC 330 RSG and those of the more metal rich LMC clusters. The solid segment of the evolutionary track in figure~\ref{combined} is the down slope of the red giant branch in which models predict the red supergiants spend $\sim$75\% of their lives when cooler than log T$_{eff}$=3.8. The spread of observed effective temperatures at a fixed luminosity is undoubtably due, in the majority, to evolution within this portion of the H-R diagram.

It is apparent that the data from both metallicity regimes fail to lie within the region delimited by down slope of the red giant branch of the appropriate metallicity. The data is offset to lower temperatures than expected. The LMC clusters show this clearly. Here we see an offset of the mean temperature of the RSG clump from the centre of the predicted band of $\Delta$logT$_{eff}$=-0.03. 

Such an offset deserves discussion. The resulting temperature of the RSG population is dependent on a large number of physical inputs to the evolutionary model. Of importance is the treatment of opacity, in particular molecular opacity, input physics such as the presence of convective envelope undershoot, the mixing length parameter $\alpha_{P}$ and the metallicity. Stothers and Chin (\cite{sto95}) find that $\delta$log T$_{eff}$/$\delta$logZ$\sim$-0.06. To account for the observed temperature of the observed RSG in the LMC clusters we would require a [Fe/H]$\sim$0. Such a metallicity is ruled out by a vast body of evidence. Similarly, to account for the shift we would have to have a E($B$$-$$V$) 0.06 less than used here. Such a modification is not compatible with observations.

The calibration of the temperature of the RSG branch is achieved within a specific model by adjustment of the $\alpha_{P}$ parameter within standard mixing length theory. Typically, a match is sought between the model prediction of a one solar mass star and that observed for the Sun. Within the model the value for $\alpha_{P}$ is generally assumed to be constant with metallicity and mass. The resulting T$_{eff}$ of the RSG is also dependent on the treatment of opacity, in particular molecular opacity and input physics such as the presence of convective undershoot. $\alpha_{P}$ is a free parameter with in each model hence its value is potentially a resting place for the effects of assumptions and uncertainties within the particular model.

To investigate the model dependence of $\alpha_{P}$ we show the models of Schaerer et al (\cite{sch93}) and Charbonnel et al. (\cite{cha93}) (hereafter the Geneva group). We see the two models differ by up to 300K in the location of the RSG region. The Geneva models offer a better match to the observational data. Both the Geneva group and the Fagotto et al. models possess ostensibly the same input physics (same $\alpha_{P}$, opacities etc.) and differ only slightly in terms of the degree of convective core overshoot and envelope undershoot (Fagotto et al. use a slightly higher degree of convective core overshoot and implement envelope undershoot). Given the inhomogeneities in the resulting RSG temperatures from various models it is not clear that we can discuss the observed offset in temperature of the RSG population in terms of an absolute numerical value for $\alpha_{P}$. Hence we discuss $\alpha_{P}$ as a free parameter within the context of the evolutionary models of Fagotto et al (\cite{fag94}) and the Geneva group respectively.

\begin{figure}
\plotone{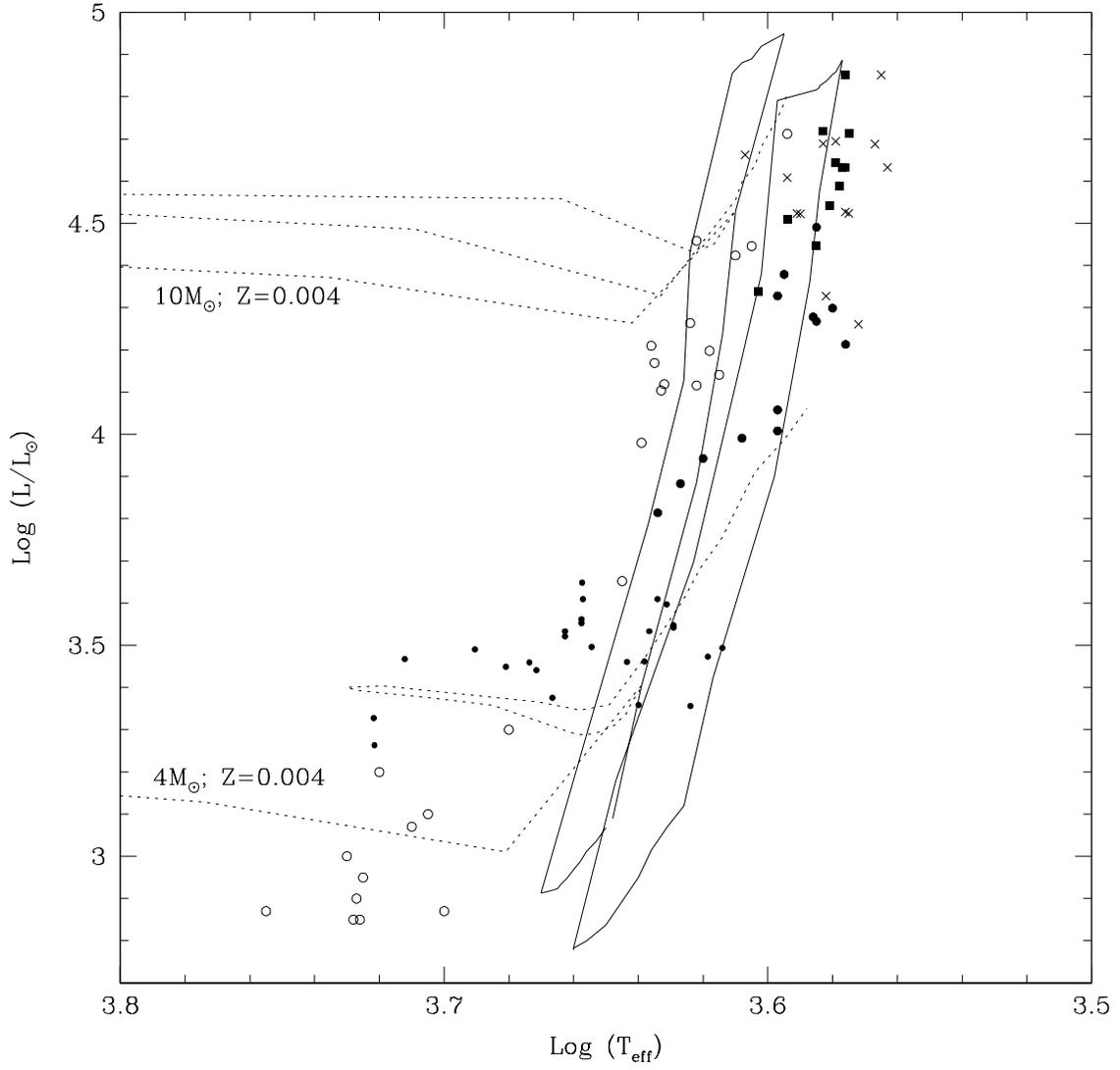}
\caption{H-R diagram showing evolutionary tracks from the models of the Geneva group for Z=0.004 (dashed lines) at 10 and 4M$_{\odot}$. Symbols are as in figure~\ref{combined}.}
\label{combinedgeneva}
\end{figure}

Firstly, we can explore the metallicity dependence of $\alpha_{P}$. The models of Fagotto et al. (\cite{fag94}) and the Geneva group are of appropriate metallicity for the LMC and SMC. These models show a difference in temperature of 0.03 dex between the centres of the respective shaded regions in figures~\ref{combined} and~\ref{combinedgeneva} at a given luminosity. The data shows a similar separation, 0.02$\pm$0.01 dex. As the models incorporate a constant value of $\alpha_{P}$ with metallicity and predict the observed {\it difference} in average T$_{eff}$ between LMC and SMC, we conclude that there is no evidence for a dependence of $\alpha_{P}$ on metallicity. Previous studies have focussed on globular clusters of considerably lower stellar mass and metallicity. Our finding is in line with the study of Pedersen et al. (\cite{ped90}) but at odds with the conclusions of Salaris \& Cassisi (\cite{sal96}). We consider the conclusion of Salaris \& Cassisi to be largely an over interpretation of the available data.

On physical grouds one might expect that $\alpha_{P}$ be a function of stellar mass. The stellar envelope of a high mass star is seen to be considerably different to that of a solar mass star. For example, the magnitude of micro-turbulence, indicative of bulk motions in the outer envelope, is seen to be a strong function of stellar mass. To investigate a trend with mass we have included in figure~\ref{combined} and~\ref{combinedgeneva} the intermediate age cluster NGC 458 in the SMC which is a useful comparison to NGC 330. We have used the $BV$ photometry of Arp (\cite{arp59}) and Walker (\cite{wal87}), and the ($B$$-$$V$) to T$_{eff}$ relation of Bessell et al. We have taken the reddening to the cluster as 0.03 (Arp \cite{arp59}). The resultant effective temparatures are systematically {\it hotter} than the predictions of both models. It should be noted that there is considerable uncertainty in the derived results for NGC 458. Stars in common between Walker and Arp show a systematic difference of $\sim$0.15 in ($B$$-$$V$). This corresponds to logT$_{eff}$ $\sim$0.02 cooler in the case of Walker's photometry. This improves upon the discrepancy but does not remove it. In addition, the reddening of E($B$$-$$V$)=0.03 is minimal; additional reddening would lead to hotter effective temperatures for the cool giants. Also shown is the position of the red giant population in the LMC cluster, NGC 1850 (from $VI$ photometry of Sebo and Wood \cite{seb94}). This data appears to be in reasonable agreement with the Fagotto et al (\cite{fag94}) model predictions, although it contains a large dispersion.

Our data shows that the standard evolutionary models do not adequately predict the temperatures of RSG and red giant populations in the range of 5-15$M_{\odot}$. In both the models of Fagotto et al (\cite{fag94}) and the Geneva group we see substaintial evidence for a mass dependance of $\alpha_{P}$. The available data indicates a trend for decreasing $\alpha_{P}$ with increasing stellar mass. Clarification of this trend will require further study at the masses associated with NGC 458 and NGC 1850.

\section{Summary} 
We have presented infra-red photometry in the $J$ and $K$ bands for a sample of red supergiants within four young clusters in the Magellanic Clouds. Such measurements provide a valuable check on predictions of stellar evolutionary models at low metallicity. We have seen that standard models of Fagotto et al (\cite{fag94}) and the Geneva group, both with similar input physics, differ markedly in the temperature they predict for the RSG population. Such differences likely arise from the treatment of the $\alpha_{P}$ parameter used in standard mixing length theory. As a free parameter $\alpha_{P}$ also typically contains the effects of assumptions and uncertainties within a given evolutionary model. 

For this reason, discussion of $\alpha_{P}$ in a quantitative manner can not be made without reference to a specific model. Qualitatively then, both models show evidence for a mass dependence of $\alpha_{P}$, specifically that $\alpha_{P}$ decreases with increasing stellar mass. We do not see evidence for a metallicity dependence of $\alpha_{P}$ within our data.

\begin{acknowledgements} 
SCK acknowledges the support of an APA scholarship.
\end{acknowledgements}

\clearpage
 
\begin{deluxetable}{lcccccccc}
\footnotesize
\tablecaption{NGC 330\label{tab1}}
\tablewidth{0pt}
\tablehead{
\colhead{Star} & \colhead{$K$}   & \colhead{$J-K$}   & \colhead{$V$} & 
\colhead{$($V$$-$$K$)_{0}$}  & \colhead{$($J$$-$$K$)_{0}$} & \colhead{${\rm BC}_{K}$} & 
\colhead{Log$T_{eff}$\tablenotemark{a}}     & \colhead{Log(L/L$_{\odot})$\tablenotemark{b}}
} 
\startdata
  A7 &  9.26 &  1.21 & 13.08 & 3.59 & 1.16 & 2.58& 3.594& 4.702\nl
  A14 & 10.01 &  1.02 & 13.61 & 3.37 & 0.98 & 2.51& 3.605& 4.433\nl
  B40 & 10.08 &  0.95 & 13.42 & 3.11 & 0.90 & 2.40& 3.622& 4.450\nl
  A6 & 10.09 &  0.98 & 13.62 & 3.30 & 0.93 & 2.48& 3.610& 4.414\nl
  A57 & 10.58 &  0.99 & 13.89 & 3.08 & 0.94 & 2.38& 3.624& 4.255\nl
  A27 & 10.71 &  1.03 & 14.11 & 3.17 & 0.98 & 2.42& 3.618& 4.187\nl
 A46 & 10.80 &  0.85 & 13.93 & 2.90 & 0.80 & 2.30& 3.636& 4.198\nl
 A45 & 10.89 &  0.80 & 13.04 & 2.92 & 0.75 & 2.31& 3.635& 4.160\nl
 A52 & 10.83 &  0.88 & 14.28 & 3.22 & 0.83 & 2.44& 3.615& 4.132\nl
 A42 & 11.00 &  0.92 & 14.21 & 2.95 & 0.87 & 2.33& 3.632& 4.110\nl
 A9 & 10.94 &  0.92 & 14.28 & 3.11 & 0.87 & 2.40& 3.622& 4.105\nl
 B42 & 11.05 &  0.86 & 14.21 & 2.93 & 0.81 & 2.32& 3.633& 4.093\nl
 B31 & 11.39 &  0.96 & 14.48 & 2.86 & 0.91 & 2.28& 3.639& 3.971\nl
 B3 & 12.25 &  0.87 & 15.26 & 2.78 & 0.82 & 2.25& 3.645& 3.642\nl

\enddata

\tablenotetext{a}{Estimated uncertainty $\pm$0.007}
\tablenotetext{b}{Estimated uncertainty $\pm$0.010}

\end{deluxetable}
 
\begin{deluxetable}{lcccccccc}
\footnotesize
\tablecaption{NGC 1818\label{tab2}}
\tablewidth{0pt}
\tablehead{
\colhead{Star} & \colhead{$K$}   & \colhead{$J-K$}   & \colhead{$V$} & 
\colhead{$($V$$-$$K$)_{0}$}  & \colhead{$($J$$-$$K$)_{0}$} & \colhead{${\rm BC}_{K}$} & 
\colhead{Log$T_{eff}$\tablenotemark{a}}     & \colhead{Log(L/L$_{\odot})$\tablenotemark{b}}
} 
\startdata
  B26&   9.36 &  1.08 & 13.35 & 3.76 & 1.03&  2.64& 3.585& 4.482\nl
  A18&   9.92 &  1.01 & 13.90 & 3.75 & 0.96&  2.63& 3.585& 4.259\nl
  C12&   9.70 &  1.06 & 13.50 & 3.57 & 1.01&  2.57& 3.595& 4.372\nl
  A15&  10.00 &  1.05 & 14.19 & 3.96 & 1.00&  2.69& 3.576& 4.203\nl
  A11&   9.81 &  0.99 & 13.89 & 3.85 & 0.94&  2.67& 3.580& 4.289\nl
  B12&   9.84 &  1.03 & 13.60 & 3.53 & 0.98&  2.56& 3.597& 4.321\nl
  A75&   9.90 &  1.02 & 13.87 & 3.74 & 0.97&  2.63& 3.586& 4.268\nl
  A93&  10.64 &  0.88 & 14.40 & 3.53 & 0.83&  2.56& 3.597& 4.001\nl
 A95&  10.52 &  1.07 & 14.27 & 3.52 & 1.02&  2.55& 3.597& 4.050\nl
 A77&  10.76 &  1.02 & 14.31 & 3.32 & 0.97&  2.48& 3.608& 3.984\nl
 B22&  10.96 &  0.91 & 14.32 & 3.13 & 0.86&  2.40& 3.620& 3.934\nl
 B1&  11.16 &  0.89 & 14.40 & 3.01 & 0.84&  2.36& 3.627& 3.873\nl
 C7&  11.38 &  0.79 & 14.52 & 2.91 & 0.74&  2.31& 3.634& 3.803\nl
\enddata

\tablenotetext{a}{Estimated uncertainty $\pm$0.007}
\tablenotetext{b}{Estimated uncertainty $\pm$0.010}

\end{deluxetable}
\begin{deluxetable}{lcccccccc}
\footnotesize
\tablecaption{NGC 2004\label{tab3}}
\tablewidth{0pt}
\tablehead{
\colhead{Star} & \colhead{$K$}   & \colhead{$J-K$}   & \colhead{$V$} & 
\colhead{$($V$$-$$K$)_{0}$}  & \colhead{$($J$$-$$K$)_{0}$} & \colhead{${\rm BC}_{K}$} & 
\colhead{Log$T_{eff}$\tablenotemark{a}}     & \colhead{Log(L/L$_{\odot})$\tablenotemark{b}}
} 
\startdata
  C7 &   8.44  &   0.86 & 12.60 & 3.93 & 0.81 & 2.69 &  3.577 &  4.830\nl
  A54&   8.80  &   1.03 & 13.25 & 4.22 & 0.98 & 2.75 & 3.567 &  4.658\nl
  A22&   8.78  &   1.16 & 13.33 & 4.32 & 1.11 & 2.78 & 3.563 &  4.657\nl
  C14&   9.15  &   0.97 & 13.21 & 3.83 & 0.92 & 2.66 & 3.581 &  4.556\nl
  B31&   8.96  &   0.90 & 13.12 & 3.93 & 0.85 & 2.69  & 3.577 &  4.622\nl
  B5 &   9.21  &   0.86 & 13.42 & 3.98 & 0.81 & 2.70  & 3.575  & 4.517\nl
  C19&   8.95  &   0.86 & 13.15 & 3.97 & 0.81 & 2.70 &  3.576 &  4.622\nl
 A64 &   9.64  &   0.95 & 13.19 & 3.31 & 0.90 & 2.48 &  3.609 &  4.433\nl
 B23 &   9.47  &   0.89 & 13.45 & 3.75 & 0.84 & 2.63 &  3.585  & 4.439\nl
 B45 &   8.84  &   0.98 & 13.45 & 4.38 & 0.93 & 2.79  & 3.561 &  4.627\nl
 A51 &   8.81  &   1.12 & 13.06 & 3.96 & 1.08 & 2.67 & 3.580 & 4.676\nl

\enddata

\tablenotetext{a}{Estimated uncertainty $\pm$0.007}
\tablenotetext{b}{Estimated uncertainty $\pm$0.010}

\end{deluxetable}
\begin{deluxetable}{lcccccccc}
\footnotesize
\tablecaption{NGC 2100\label{tab4}}
\tablewidth{0pt}
\tablehead{
\colhead{Star} & \colhead{$K$}   & \colhead{$J-K$}   & \colhead{$V$} & 
\colhead{$($V$$-$$K$)_{0}$}  & \colhead{$($J$$-$$K$)_{0}$} & \colhead{${\rm BC}_{K}$} & 
\colhead{Log$T_{eff}$\tablenotemark{a}}     & \colhead{Log(L/L$_{\odot})$\tablenotemark{b}}
} 
\startdata
  B40 & 8.71 &  1.42 & 13.47 & 4.06 & 1.28 & 2.72&  3.572&  4.709\nl
  C2 &  9.02 &  1.29 & 13.16 & 3.44 & 1.15 & 2.53&  3.601&  4.661\nl
  B4 &  9.00 &  1.37 & 13.75 & 4.14 & 1.23 & 2.84&  3.572&  4.688\nl
  A127& 9.06 &  1.34 & 13.39 & 3.63 & 1.20 & 2.59&  3.591&  4.619\nl
 C32 &  9.05 &  1.34 & 14.09 & 4.18 & 1.20 & 2.85&  3.571&  4.632\nl
 B47 &  9.07 &  1.34 & 13.45 & 3.68 & 1.20 & 2.61&  3.589&  4.608\nl
 A74 &  9.23 &  1.30 & 13.45 & 3.52 & 1.16 & 2.56&  3.597&  4.566\nl
 A81 &  9.38 &  1.23 & 13.92 & 3.84 & 1.09 & 2.66&  3.581&  4.462\nl
 C12 &  9.51 &  1.16 & 13.70 & 3.49 & 1.02 & 2.55&  3.598&  4.458\nl
 B22 &  9.46 &  1.19 & 13.72 & 3.56 & 1.05 & 2.57&  3.595&  4.468\nl
 B17 &  9.32 &  1.31 & 13.95 & 3.93 & 1.17 & 2.69&  3.577&  4.477\nl
 B46 &  9.88 &  1.19 & 14.31 & 3.73 & 1.05 & 2.63&  3.586&  4.277\nl
 C34 & 10.00 &  1.12 & 14.67 & 3.97 & 0.98 & 2.70&  3.576&  4.201\nl

\enddata

\tablenotetext{a}{Estimated uncertainty $\pm$0.007}
\tablenotetext{b}{Estimated uncertainty $\pm$0.010}

\end{deluxetable}


\end{document}